\documentstyle[12pt,citesort]{article}

\catcode`\@=11

\newcount\hour
\newcount\minute
\newtoks\amorpm
\hour=\time\divide\hour by 60
\minute=\time{\multiply\hour by 60 \global\advance\minute by-\hour}
\edef\standardtime{{\ifnum\hour<12 \global\amorpm={am}%
        \else\global\amorpm={pm}\advance\hour by-12 \fi
        \ifnum\hour=0 \hour=12 \fi
        \number\hour:\ifnum\minute<10
        0\fi\number\minute\the\amorpm}}
\edef\militarytime{\number\hour:\ifnum\minute<10
0\fi\number\minute}

\def\draftlabel#1{{\@bsphack\if@filesw {\let\thepage\relax
   \xdef\@gtempa{\write\@auxout{\string
      \newlabel{#1}{{\@currentlabel}{\thepage}}}}}\@gtempa
   \if@nobreak \ifvmode\nobreak\fi\fi\fi\@esphack}
        \gdef\@eqnlabel{#1}}
\def\@eqnlabel{}
\def\@vacuum{}
\def\marginnote#1{}
\def\draftmarginnote#1{\marginpar{\raggedright\scriptsize\tt#1}}
\def\draft{
        \pagestyle{plain}
        \overfullrule=2pt
        \oddsidemargin -.5truein
        \def\@oddhead{\sl \phantom{\today\quad\militarytime} \hfil
        \smash{\Large\sl DRAFT} \hfil \today\quad\militarytime}
        \let\@evenhead\@oddhead
        \let\label=\draftlabel
        \let\marginnote=\draftmarginnote
        \def\ps@empty{\let\@mkboth\@gobbletwo
        \def\@oddfoot{\hfil \smash{\Large\sl DRAFT} \hfil}
        \let\@evenfoot\@oddhead}
        \def\@eqnnum{(\theequation)\rlap{\kern\marginparsep\tt\@eqnlabel}%
        \global\let\@eqnlabel\@vacuum}  }

\renewcommand{\theequation}{\thesection.\arabic{equation}}
\renewcommand{\thefootnote}{\fnsymbol{footnote}}

\newcommand{\newsection}{    
\setcounter{equation}{0}
\section}
\def\appendix#1{
  \addtocounter{section}{1}
  \setcounter{equation}{0}
  \renewcommand{\thesection}{\Alph{section}}
  \section*{Appendix \thesection\protect\indent \parbox[t]{11.15cm} {#1} }
  \addcontentsline{toc}{section}{Appendix \thesection\ \ \ #1}
  }

\def\apr{{{\rm A}^\prime}}
\def\bpr{{{\rm B}^\prime}}
\def\cpr{{{\rm C}^\prime}}

\def\epr{{{\rm E}^\prime}}

\def\sca{{\scriptscriptstyle{\cal  A}}}
\def\scb{{\scriptscriptstyle{\cal  B}}}

\def\aA{{\rm A}}
\def\aB{{\rm B}}
\def\aC{{\rm C}}
\def\aE{{\rm E}}

\def\aA{{A}}
\def\aB{{B}}
\def\aC{{C}}
\def\aE{{E}}

\def\N{{\cal N}}
\def\Csp{C^\prime}

\def\x'{\hbox{\'x}}
\def\y'{\hbox{\'y}}

\def\X'{\hbox{\'X}}

\def\al{{\perp}}

\def\alpr{i}
\def\bepr{j}

\def\vm{{\mu}}
\def\vn{{\nu}}

\def\ssmP{{\scriptscriptstyle P}}
\def\ssmK{{\scriptscriptstyle K}}
\def\ssmD{{\scriptscriptstyle D}}
\def\ssmQ{{\scriptscriptstyle Q}}
\def\ssmS{{\scriptscriptstyle S}}

\def\sfa{{\sf a}}
\def\sfb{{\sf b}}
\def\sfc{{\sf c}}
\def\sfd{{\sf d}}

\def\dsfa{{\dot{\sf a}}}
\def\dsfb{{\dot{\sf b}}}
\def\dsfc{{\dot{\sf c}}}
\def\dsfd{{\dot{\sf d}}}

\def\PP{{\cal P}}

\def\x'{\mathaccent 19 x}
\def\y'{\mathaccent 19 y}
\def\n'{\mathaccent 19 n}
\def\u'{\mathaccent 19 u}

\def\X'{\mathaccent 19 X}
\def\Y'{\mathaccent 19 Y}
\def\Z'{\mathaccent 19 Z}

\def\et'{\mathaccent 19 \eta}
\def\th'{\mathaccent 19 \theta}
\def\lam'{\mathaccent 19 \lambda}
\def\varet'{\mathaccent 19 \vartheta}
\def\rh'{\mathaccent 19 \rho}
\def\ph'{\mathaccent 19 \phi}
\def\xb'{\mathaccent 19 {\bar{x}}}

\def \bi{\bibitem}
\def \la {\label}

\def\nline{\,\nabla\kern -0.7em\raise0.2ex\hbox{/}\,\,}
\def\yline{\,y\kern -0.47em /}
\def\aline{\,a\kern -0.49em /}
\def\parline{\,\partial\kern -0.55em /\,\,}

\def\be{\begin{equation}}
\def\ee{\end{equation}}

\def \ci {\cite}

\begin{document}

\begin{titlepage}
\begin{flushright}
FIAN/TD/00-23
\\
OHSTPY-HEP-T-00-031
\\
hep-th/0012026\\
\end{flushright}
\vspace{.5cm}

\begin{center}
{\LARGE
On manifest $SU(4)$ invariant 
superstring \\ [.1cm] action
 in AdS$_5 \times $S$^5$
 }\\[.2cm]
\vspace{1.1cm}
{\large R.R. Metsaev${}^{{\rm a,b,}}$\footnote{\
E-mail: metsaev@lpi.ru, metsaev@pacific.mps.ohio-state.edu}
}

\vspace{18pt}
 ${}^{{\rm a\ }}${\it
 Department of Physics,
The Ohio State University  \\
Columbus, OH 43210-1106, USA\\
}

\vspace{6pt}

${}^{{\rm b\ }}${\it
Department of Theoretical Physics, P.N. Lebedev Physical
Institute,\\ Leninsky prospect 53,  Moscow 117924, Russia
}
\end{center}

\vspace{2cm}

\begin{abstract}

We discuss manifestly  $SL(2,C) \times SU(4)$ and $\kappa$ 
invariant superstring action in $AdS_5 \times S^5$ background 
in the framework of Green-Schwarz  formulation.
The action is formulated in terms of 16 Poincar\'e 
fermionic coordinates which through AdS/CFT correspondence 
should represent    $\N=4$ 
SYM superspace 
and 16 superconformal fermionic coordinates. 
The action is also manifestly invariant with respect to
the  usual 
$\N=4$ Poincar\'e superalgebra transformations.
$\kappa$-symmetry gauge 
fixing and the derivation of light-cone gauge action is 
simplified.
\end{abstract}

\end{titlepage}
\setcounter{page}{1}
\renewcommand{\thefootnote}{\arabic{footnote}}
\setcounter{footnote}{0}

\def \adss {$AdS_5 \times S^5$\ }

\newsection{Introduction and summary of result}
\def \N {{\cal N}}

Motivated by  the conjectured duality between the string theory
and ${\cal N}=4$, $4d$ SYM theory \cite{mal} the Green-Schwarz
formulation of strings propagating in $AdS_5\times S^5$ was
suggested in \cite{MT} (for  further developments see
\cite{krr,k,pes,kr}).   Alternative approaches based on exploiting
twistor like  variables can be found in \cite{GK,ber,RS}.  
Recently
some progress in understanding light-cone gauge  formulation has
been achieved in \cite{MT2,MTT}. 
The action found in \cite{MT2,MTT} in the particle limit
describes  massless states of $IIB$ supergravity and therefore 
it can
be used to discuss sector of short fundamental strings. 
Alternative gauge fixed action found in \cite{k,pes,kr}, 
so called S-gauge action, describes sector of long strings
\cite{kt,dgt};  this action is suitable 
 for discussion of various BPS
configurations of $AdS$ superstrings. 

The original action of superstrings in $AdS_5\times S^5$ 
\cite{MT} being realized as sigma model on
coset superspace $PSU(2,2|4)/SO(4,1)\times SO(5)$ has manifest
symmetry with respect to $SO(4,1)\times SO(5)$.
An  interesting phenomenon  discovered in
\cite{kr,MT2} is that after imposing $\kappa$-symmetry gauge the
resulting action can be cast into manifestly $SU(4)$  invariant 
form. This fact while  being expected for the 
 bosonic part
of action, surprisingly 
 turns out to be also a  feature of the fermionic sector 
of the  GS action. 

Here we would like to demonstrate 
that that not only gauge fixed actions have $SU(4)$
invariant formulation but the original $\kappa$-symmetric  GS 
action
\cite{MT}  can be put  into a manifestly
$SU(4)$ invariant form.
To derive $\kappa$ and manifestly $SU(4)$
invariant superstring action in the framework of GS formulation
is desirable for several reasons.
Such  $\kappa$-invariant action can be used to study
interaction of superstring with $D3$ brane. As is well known,
$\kappa$-symmetry plays
defining role and can be used 
to fix interactions of superstring ending 
on $D3$ brane. 
The formulation we 
develop is  based on splitting fermionic coordinates into $\theta$ 
and $\eta$. The variables $\theta$ represent the  odd part 
of ${\cal N}=4$ Poincar\'e superalgebra superspace and they can 
be responsible for the description of boundary $\N=4$ SYM theory. 
The $\eta$ are superconformal fermionic coordinates and they 
have nonlinear dynamics (even light-cone action involves terms 
of fourth degree in $\eta$). Because $\N=4$ SYM theory respects 
manifest $SU(4)$ invariance it is desirable to put superstring
action 
into manifestly  $SU(4)$ invariant form from the very beginning, 
i.e. at the level of $\kappa$-invariant action.

The superstring Lagrangian is formulated in terms of 10 bosonic coordinates
$(x^a,Z^M)$, ($x^a$ are 
4 coordinates along boundary directions, while $Z^M$
 represent  one $AdS$ radial coordinate 
and five coordinates of $S^5$) and 32 two fermionic
coordinates $(\theta_i^\sfa,\theta^{\dsfa i},\eta^i_\sfa,\eta_i^\dsfa)$
which transform in (anti)fundamental
representations of $SU(4)$ with respect to indices $i$ and one-half
representations of $sl(2,C)$ with respect to indices $\sfa,\dsfa$.
The Lagrangian is given by the sum of kinetic ${\cal L}_{kin}$ and
Wess-Zumino terms

\be\label{lag0}
{\cal L} = {\cal L}_{kin} +{\cal L}_{WZ}\,,
\ee

\be\label{lag0kin}
{\cal L}_{kin} = -\frac{1}{2}\sqrt{g}g^{\vm\vn}
(\hat{L}_\vm ^{\sfa\dsfb} \hat{L}_{\vn \sfa\dsfb} 
+\frac{1}{Z^2}D_\vm Z^M D_\vn Z^M)\,,
\ee

\be\label{lwz1}
f{\cal L}_{WZ}= 
{\rm i}|Z|^{-2} d\theta^{\dsfa i}Z_{ij} d\theta_{\dsfa}^j
+\int_0^1 \!\! dt
\Bigl[2|Z|^{-1}\hat{L}^{\sfa \dsfb}
\eta_\sfa^i Z_{ij}{\sf L}_{\ssmQ \dsfb}^j
+{\rm i}\eta^{\sfa i}DZ_{ij}{\sf L}_{\ssmS \sfa}^j\Bigr]
+h.c.\,,
\ee
where the $2d$ metric $g_{\vm\vn}$, $\vm,\vn=0,1$ has signature 
$(-,+)$, $g\equiv -det g_{\vm\vn}$.
We use  the notation 

\be
Z_{ij} \equiv \rho_{ij}^M Z^M\,,
\qquad
|Z|^2 =Z^MZ^M\,,
\qquad
M = 1,\ldots ,6
\ee

\be
f=2\sqrt{2}\,.
\ee
The `covariant' derivative $DZ^M$ is defined by

\be\label{covder1}
DZ^M = dZ^M 
-\frac{1}{2}\Bigl(
d\theta_i^\sfa (\rho^N\bar{\rho}^M)^i{}_j \eta_\sfa^j
+\eta_i^\dsfa (\rho^M\bar{\rho}^N)^i{}_j d\theta_\dsfa^j
+{\rm i}{\sf L}_\ssmP^{\sfa\dsfb}
\eta_{\dsfb i}(\rho^{MN})^i{}_j\eta_\sfa^j\Bigr)Z^N\,,
\ee
where the $\rho^M =(\rho^M)^{ij}$, $\bar{\rho}^M = \rho_{ij}^M$ 
are $4\times 4$
$SO(6)$ $\gamma$ matrices in chiral representation,
and $\rho^{MN} \equiv \rho^{[M}\bar{\rho}^{N]}$,
$\bar{\rho}^M= \rho^{M \dagger}$ (see Appendix A).
The $\hat{L}^{\sfa\dsfb}$ are bosonic left invariant Cartan 1-forms 
while ${\sf L}_{\ssmQ i}^\sfa$ and ${\sf L}_{\ssmS \sfa }^i$ are
the fermionic ones. They are given by

\be
\hat{L}^{\sfa\dsfb } = L_\ssmP^{\sfa\dsfb}
-\frac{1}{2}L_{\ssmK}^{\sfa\dsfb}\,,
\qquad
L_\ssmP^{\sfa\dsfb} = |Z|^{-1} {\sf L}_\ssmP^{\sfa\dsfb}\,,
\qquad
L_\ssmK^{\sfa\dsfb} = |Z|{\sf L}_\ssmK^{\sfa\dsfb}\,,
\ee

\be
{\sf L}_{\ssmP}^{\sfa\dsfb}
=dx^{\sfa\dsfb}-\frac{\rm i}{2}(\theta_i^\sfa d\theta^{\dsfb i}
+\theta^{\dsfb i}d\theta^\sfa_i)\,,
\ee

\be\label{LK}
{\sf L}_\ssmK^{\sfa\dsfb}
= \frac{\rm i}{2}\eta^{\sfa i}d\eta_i^\dsfb
+\frac{\rm i}{12}(d\theta_i^\sfc
-\frac{\rm i}{4}{\sf L}_{\ssmP}^{\sfc\dsfd}\eta_{\dsfd i})
(5\eta_{\sfc}^j\eta^{\sfa i}
-\eta_{\sfc}^i\eta^{\sfa j}
-\eta^{\sfd i}\eta_\sfd^j\delta_\sfc^\sfa)\eta_j^\dsfb
+h.c.\ \,,
\ee

\be
{\sf L}_\ssmQ^{\dsfa i}
=d\theta^{\dsfa i} -{\rm i}\eta_\sfb^i {\sf L}_{\ssmP}^{\sfb\dsfa}\,,
\qquad
{\sf L}_{\ssmQ i}^{\sfa}
= d\theta_i^\sfa  -{\rm i}{\sf L}_{\ssmP}^{\sfa\dsfb}\eta_{\dsfb i}\,,
\ee

\be
{\sf L}_\ssmS^{\sfa i}
=d\eta^{\sfa i}+ \frac{1}{4}(d\theta_j^\sfc  
-\frac{\rm i}{3}{\sf L}_{\ssmP}^{\sfc\dsfd}\eta_{\dsfd j})
(\eta_{\sfc}^j\eta^{\sfa i}
-3\eta_{\sfc}^i\eta^{\sfa j}
+\eta^{\sfd j}\eta_\sfd^i\delta_\sfc^\sfa)
+\frac{1}{2}(d\theta^{\dsfc i} 
-\frac{\rm i}{3}\eta_\sfb^i {\sf L}_{\ssmP}^{\sfb\dsfc})
\eta_{\dsfc j}\eta^{\sfa j}\,.
\ee
Their world-sheet projections are 
defined as usual by $L = d\sigma^\vm L_\vm$. Hermitean conjugation in 
(\ref{LK}) should be supplemented by $\sfa \leftrightarrow \sfb$.
The $x^{\sfa\dsfb}$ is expressible 
in terms of the coordinates $x^a$ as

\be
x^{\sfa\dsfb}=\frac{1}{\sqrt{2}}(\sigma^a)^{\sfa\dsfb}x^a\,,
\ee 
where 
$(\sigma^a)^{\sfa\dsfb}$ are $2\times 2$ 
$SO(3,1)$ $\gamma$ matrices 
in chiral representation (see Appendix A).
Note that 
in expressions for $\hat{L}^{\sfa\dsfb}$, ${\sf L}_{\ssmQ \dsfa}^i$, 
${\sf L}_{\ssmS \sfa}^i$, $DZ^M$ in 
${\cal L}_{WZ}$ (\ref{lwz1}) the $\eta$ should be shifted 
$\eta \rightarrow t\eta$. 
In (\ref{lwz1})
the products of 1-forms, say $L^1$ and $L^2$,
should read as $\epsilon^{\vm\vn} L_\vm^1 L_\vn^2$.
The above action corresponds to the ``$4+6$'' choice of 
conformally flat 
coordinates in $AdS_5 \times S^5$ space

\be\label{concoo}
ds^2 =\frac{1}{Z^2}(dx^a dx^a + dZ^M dZ^M)\,.
\ee
Here and below we set the radius of $AdS_5$ and $S^5$ to 1. 
We finish this discussion with few remarks.

(i) The Lagrangian (\ref{lag0}) is manifestly invariant 
with respect to $SL(2,C)\times SU(4)$.

(ii) The dependence on boundary coordinates is through 
their derivatives $dx^a$, i.e. the invariance 
with respect to boundary Poincar\'e translations is manifest.

(iii) The Lagrangian depends on Poincar\'e fermionic coordinates 
$\theta$ either through their derivatives $d\theta$ or fermionic 1-forms 
${\sf L}_\ssmP^{\sfa\dsfb}$. Because both of these quantities are
invariant with respect to $\N=4$ Poincar\'e supersymmetries

\be 
\delta \theta^{\dsfa i} = \epsilon^{\dsfa i}\,,
\qquad
\delta \theta_i^\sfa = \epsilon_i^\sfa\,,
\qquad
\delta x^{\sfa\dsfb} = \frac{\rm i}{2}(\epsilon_i^\sfa \theta^{\dsfb i}
+\epsilon^{\dsfb i}\theta_i^\sfa)\,,
\ee
the Lagrangian is also manifestly invariant with respect to 
$\N=4$ Poincar\'e supersymmetry.

(iv) The Lagrangian is manifestly invariant with respect 
to dilatation symmetries which are realized as follows

\be
\delta x^{\sfa\dsfb} = e^\lambda x^{\sfa\dsfb}\,,
\quad
\delta Z^M = e^\lambda Z^M\,,
\quad
\delta\phi = -\lambda\,,
\quad
\delta \theta = e^{\lambda/2}\theta\,,
\quad
\delta \eta = e^{-\lambda/2}\eta\,.
\ee

(v) The Lagrangian involves terms of 4th degree in $\theta$ and 
8th degree in $\eta$. Maximal degree of fermionic 
coordinates appearing in Lagrangian is equal to 12.

(vi) The $S$ gauge fixed action is obtainable 
by setting $\eta=0$. In this gauge
we have $L_\ssmK^{\sfa\dsfb}=0$, ${\sf L}_{\ssmS}^{\sfa i}=0$,
$DZ^M = dZ^M$. With these relations we get immediately the $S$
gauged action \cite{pes,kr,MT2} in the form given in \cite{MT2} 
(see formulas (C.12), (C.14) in Ref. \cite{MT2}).

(vii) Light-cone gauge fixed action of \cite{MT2} 
is obtainable by setting 

\be\label{lcg1}
\theta_i^2 = \theta_{1i} = 
\theta^{\dot{2}i} = \theta_{\dot{1}}^i=
\eta^{2i} = \eta_1^i = 
\eta^{\dot{2}}_i = \eta_{\dot{1} i}=0\,,
\ee
\be\label{lcg2}
\theta^{\dot{1}i} = -\theta_{\dot{2}}^i =  - \theta^i\,,
\qquad
\theta^{1}_i = \theta_{2i} = -\theta_i\,,
\qquad
\eta^{1i} = \eta_2^i = \eta^i\,,
\qquad
\eta^{\dot{1}}_i = -\eta_{\dot{2} i} = \eta_i\,.
\ee
Note that 
due to index  raising and lowering rules (\ref{rlrul}) 
and Hermitean conjugation rules 

\be
(\theta_i^\sfa)^\dagger = \theta^{\dsfa i}\,,
\quad
(\theta_{\sfa i})^\dagger = -\theta_\dsfa^i\,,
\quad
(\eta^{\sfa i})^\dagger = \eta_i^\dsfa\,,
\quad
(\eta_{\sfa}^i)^\dagger = -\eta_{\dsfa i}\,,
\ee
only one quarter of the  relations (\ref{lcg1}),(\ref{lcg2}) 
are independent. 
In light-cone gauge all bilinear expressions for fermions 
with contracted  $sl(2)$ indices are 
equal to zero, e.g., 

\be
\theta_i^\sfa \theta_{\sfa j}=0\,,
\qquad
\eta^{i\sfa} \eta^j_\sfa=0\,,
\qquad
\theta_i^\sfa \eta_\sfa^j=0\,.
\ee
 Note that due to this rule the first and the last terms in 
${\cal L}_{WZ}$ (\ref{lwz1}) are equal to zero.  Taking into account 
that in light-cone gauge 
$L_\ssmK^{2\dot{2}}=
L_\ssmK^{1\dot{2}}=
L_\ssmK^{2\dot{1}}=
L_{\ssmS}^{2i}=L_{\ssmS i}^{\dot{2}}=0$ 
and the relation between  $sl(2)$ and light-cone notation

\be
x^{1\dot{1}} = x^-\,,
\qquad
x^{2\dot{2}} =  - x^+\,,
\qquad
x^{1\dot{2}} = \bar{x}\,,
\qquad
x^{2\dot{1}} = x\,,
\ee
(see Appendix A) and making field redefinitions

\be
\eta^i \rightarrow \sqrt{2} |Z|^{-1}\eta^i,
 \qquad
\eta_i \rightarrow \sqrt{2} |Z|^{-1}\eta_i, \qquad
\qquad
x^a \rightarrow - x^a\,,
\ee
we get from (\ref{lag0kin}),(\ref{lwz1}) the 
superstring kappa symmetry light-cone 
gauge fixed action found in \cite{MT2} 

$$
{\cal L}_{kin}
=
-\sqrt{g}g^{\vm\vn}Z^{-2}\Bigl[
\partial_\vm x^+ \partial_\vn x ^-
+ \partial_\vm x\partial_\vn\bar{x}
+\frac{1}{2}D_\vm Z^M D_\vn Z^M\Bigr]
$$
\begin{equation}
- \ \frac{{\rm i}}{2} \sqrt{g}g^{\vm\vn}
Z^{-2}\partial_\vm x^+
\Bigl[\theta^i\partial_\vn \theta_i
+\theta_i\partial_\vn \theta^i
+\eta^i\partial_\vn \eta_i
+\eta_i\partial_\vn \eta^i 
+{\rm i}  Z^{-2}\partial_\vn x^+(\eta^2)^2\Bigr]\ ,
\label{actkin5n}
\end{equation}
\begin{equation}\label{actwz5n}
{\cal L}_{WZ}
=\epsilon^{\vm\vn}
|Z|^{-3}\partial_\vm x^+ \eta^i \rho_{ij}^M Z^M
(\partial_\vn\theta^j-{\rm i}\sqrt{2}|Z|^{-1} \eta^j
\partial_\vn x)+h.c.\ , 
\end{equation}
where in light-cone gauge (\ref{lcg1}),(\ref{lcg2}) the covariant derivative 
(\ref{covder1}) reduces to

\be 
DZ^M = dZ^M -2 {\rm i}\eta_i (R^{M})^i{}_j\eta^j  Z^{-2} dx^+\, ,
\qquad\ \  R^M  = - \frac{1}{2} \rho^{MN} Z^N \ .
\ee 

(viii)
By applying phase space GGRT \cite{GGRT} approach based on fixing 
diffeomorphisms by $x^+=\tau$, $\PP^+ = p^+ =const$, where $\PP^+$
is a canonical momentum, the Lagrangian 
can be put into the form found in \cite{MTT}

\begin{eqnarray}
{\cal L} 
&=& \PP_\al\dot{x}_\al 
+ \PP_M \dot{Z}^M
+\frac{{\rm i}}{2}p^+(\theta^i \dot{\theta}_i
+\eta^i\dot{\eta}_i+\theta_i \dot{\theta}^i+\eta_i\dot{\eta}^i)
\nonumber\\
&-&\frac{1}{2p^+}\Bigl[ \PP_\al^2 +\PP_M^2
+|Z|^{-4}(\x'_\al^2+ \Z'_M^2)
+|Z|^{-2}(p^{+2}(\eta^2)^2 + 4{\rm i}p^+\eta R^M\eta \PP_M)\Bigr]
\nonumber\\
&+&\Bigl[\ |Z|^{-3}\eta^i \rho_{ij}^M Z^M
(\th'^j - {\rm i}\sqrt{2}|Z|^{-1} \eta^j\x')+h.c.\Bigr]
\end{eqnarray}
where $x^\al =(x,\bar{x})$, $\dot{x}=\partial_\tau x$,
$\x'\equiv \partial_\sigma x$. 
$\PP^\al =(\PP,\bar{\PP})$ and $\PP^M$ 
are canonical momenta for $(\bar{x}, x)$ and $Z^M$ respectively. 
The interesting feature of this Lagrangian is that the squares
of the $AdS_5$ and $S^5$  momenta  enter exactly 
as in  the flat space case. 
This implies that solution to phase space light-cone gauge 
equations of motion of {\it bosonic particle} 
in $AdS_5\times S^5$ takes the same form as in flat space

\be
\PP^\al  = \PP_0^\al\,,
\quad
x^\al = x^\al_0 + \frac{\PP_0^\al}{p^+} \tau\,,
\qquad
\PP^M = \PP_0^M\,,
\quad
Z^M = Z_0^M + \frac{\PP_0^M}{p^+} \tau\,,
\ee
where the variables 
carrying subscript 0 are initial data.
An intersting feature of this solution is that as compared to the one
discussed in \cite{kt} 
our solution does not involve trigonometric functions.
This simplification is due to
(i) choice of conformally flat parametrization 
of $AdS_5\times S^5$ background;
(ii) fixing diffeomorphisms by light-cone gauge \cite{MTT}.
Note that in the case of bosonic 
particle such simple solution can be also reached
by choosing covariant gauge on world line vielbein $e=|Z|^{-2}$ which
is similar to 
the conformal gauge suggested for the string in \cite{POL}.
Taking this into account we conclude that at least in particle
approximation the conformally flat coordinates (\ref{concoo}) 
are most suitable for 
analysis of $AdS$ string dynamics.

In the rest of the paper we expalain the procedure of 
derivation of the manifestly $SL(2,C)\times SU(4)$ action from the original 
superstring action \cite{MT} which has manifest 
$SO(4,1)\times SO(5)$ invariance.

\newsection{Superstring action in $so(4,1)\oplus so(5)$ and
$sl(2,C)\oplus su(4)$ bases}

Superstring Lagrangian in $AdS_5\times S^5$  \cite{MT}
in $so(4,1)\oplus so(5)$ basis of $psu(2,2|4)$ superalgebra
has the same structure as  the flat space  GS  action \cite{GS,HM}

\begin{equation}\label{action}
{\cal L}={\cal L}_{kin}
+{\cal L}_{WZ}\,, 
\qquad
{\cal L}_{WZ} = d^{-1}( {\rm i}{\cal H})    \ ,
\end{equation}
where the kinetic term  of the \adss GS  action ${\cal L}_{kin}$
and the  WZ three form ${\cal H}$
are  expressed  in terms of the Cartan 1-forms
and have  the following
form  in  the $so(4,1)\oplus so(5)$ basis \ci{MT}

\begin{equation}\label{lkin2}
{\cal L}_{kin}=-\frac{1}{2}\sqrt{g}g^{\vm\vn}
(\hat{L}_{\vm}^\aA \hat{L}_{\vn}^\aA
+L_{\vm}^\apr L_{\vn}^\apr) \ ,
\end{equation}
\begin{equation}\label{wzt}
{\cal H}=s^{IJ}\hat{L}^\aA \bar{L}^I\gamma^\aA L^J
+{\rm i}s^{IJ}L^\apr \bar{L}^J\gamma^\apr L^J  \ ,
\end{equation}
where $s^{IJ}=diag(1,-1)$, $I,J=1,2$ and 
$\bar{L}^I \equiv L^I CC^\prime$. 
The $\gamma^A$ and $\gamma^\apr$ are $SO(4,1)$ 
and $SO(5)$ Dirac $\gamma$ matrices 
respectively while $C$ and $C^\prime$ 
are appropriate charge conjugation matrices. The left invariant
bosonic Cartan 1-forms $L^\aA$, $L^\apr$ 
and fermionic ones $L^I$  are  
defined by decomposition  

\be\label{cos55}
G^{-1}dG=(G^{-1}dG)_{bos}
+L^{I\alpha \alpr}Q_{I\alpha\alpr}
\ , \ee
where $G$ is a coset representative of $PSU(2,2|4)$ group
and the restriction to  the  bosonic part is

\begin{equation}\label{cf1}
(G^{-1}dG)_{bos}
=\hat{L}^\aA\hat{P}^\aA+\frac{1}{2}\hat{L}^{AB}\hat{J}^{\aA\aB}
+L^\apr P^\apr+\frac{1}{2} L^{\apr\bpr} J^{\apr\bpr}\ .
\end{equation}
The generators $\hat{P}^\aA$, $\hat{J}^{\aA\aB}$, and 
$P^\apr$, $J^{\apr\bpr}$ are translations and rotations generators
for $AdS_5$ and
$S^5$ respectively, while $Q_I$ are fermionic generators (see Appendix B).
The remarkable property of 
$so(4,1)\oplus so(5)$ basis is that
it is this basis that allows one
to present the \adss GS action and $\kappa$ symmetry transformations
in  the  form similar to the one in the flat space (see \cite{MT}). 

Our present  goal   is to rewrite the action in
the $sl(2,C)\oplus su(4)$ basis.
To do that we shall use the   conformal algebra and $su(4)$ 
notation.
We introduce the
Poincar\'e translations $P^a$, the conformal boosts $K^a$,
the dilatation $D$, $so(3,1)$ Lorentz algebra generators 
$J^{ab}$ and $su(4)$ generators $J^i{}_j$ by relations

\be
P^a = \hat{P}^a +\hat{J}^{4a}
\ , \ \ \ \
K^a=\frac{1}{2}(-\hat{P}^a +\hat{J}^{4a})\ ,  \ \ \ \
D=-\hat{P}^4\,,
\qquad
J^{ab} = \hat{J}^{ab} \ ,  \label{dp4}
\ee
\begin{equation}\label{su4def}
J^\alpr{}_\bepr \equiv
-\frac{{\rm i}}{2}(\gamma^\apr)^\alpr{}_\bepr P^\apr
+\frac{1}{4}(\gamma^{\apr\bpr})^\alpr{}_\bepr J^{\apr\bpr}   \ .
\end{equation}
In conformal algebra notation we have the following decomposition 

\begin{equation}\label{cf2}
(G^{-1}dG)_{bos}
=L_\ssmP^aP^a+L_\ssmK^a K^a +L_\ssmD D + \frac{1}{2}L^{ab}J^{ab}
+L^\alpr{}_\bepr J^\bepr{}_\alpr      \ .
\end{equation}
Comparing this  with (\ref{cf1}) and using (\ref{dp4}),
(\ref{su4def}) we get interrelation of Cartan 1-forms in 
$so(4,1)\oplus so(5)$ and conformal bases

\be\label{lllll}
\hat{L}^a = L_\ssmP^a - \frac{1}{2}L_\ssmK^a
\ , \ \ \ \
\hat{L}^{4a}=L_\ssmP^a+\frac{1}{2}L_\ssmK^a
\ , \ \ \ \
\hat{L}^4=-L_\ssmD\ ,
\ \ \ \
L^\apr = -\frac{\rm i}{2}(\gamma^\apr)^i{}_j L^j{}_i\,,
\ee
\be
\label{lijlab}
L^i{}_j=\frac{\rm i}{2}(\gamma^\apr)^i{}_j L^\apr
-\frac{1}{4}(\gamma^{\apr\bpr})^i{}_j L^{\apr\bpr}   \ .
\ee
Using  these representations for $\hat{L}^a$, $L_\ssmD$, $L^\apr$
allows us transform kinetic term to conformal algebra notation.
In what follows we prefer use the $sl(2,C)$ notation instead of the 
$so(3,1)$ one. To this end we introduce generators in $sl(2)$ basis

\be
P_{\sfa\dsfb} 
= \frac{1}{\sqrt{2}}(\sigma^a)_{\sfa\dsfb} P^a\,,
\qquad
K_{\sfa\dsfb} 
= \frac{1}{\sqrt{2}}(\sigma^a)_{\sfa\dsfb} K^a\,,
\ee
\be
J_{\sfa\sfb}=\frac{1}{2}(\sigma^{ab})_{\sfa\sfb}J^{ab}\,,
\qquad
J_{\dsfa\dsfb}
=-\frac{1}{2}(\bar{\sigma}^{ab})_{\dsfa\dsfb}J^{ab}\,,
\ee
and corresponding bosonic $sl(2)$ Cartan 1-forms 

\be
L_\ssmP^{\sfa\dsfb} 
= \frac{1}{\sqrt{2}}(\sigma^a)^{\sfa\dsfb}L_\ssmP^a\,,
\qquad
L_\ssmK^{\sfa\dsfb} 
= \frac{1}{\sqrt{2}}(\sigma^a)^{\sfa\dsfb}L_\ssmK^a\,,
\ee
\be
L^{\sfa\sfb}=\frac{1}{2}(\sigma^{ab})^{\sfa\sfb}L^{ab}\,,
\qquad
L^{\dsfa\dsfb}
=-\frac{1}{2}(\bar{\sigma}^{ab})^{\dsfa\dsfb}L^{ab}\,.
\ee
Transformation of supergenerators and fermionic Cartan 1-forms 
to $sl(2)\oplus su(4)$ basis may be found in Appendix B. 
In the $sl(2)\oplus su(4)$  notation Cartan 1-form are obtainable 
from the decompositon

\begin{eqnarray}\label{cf4}
G^{-1}dG
&=&
L_\ssmP^{\sfa\dsfb} P_{\sfa\dsfb}+L_\ssmK^{\sfa\dsfb} K_{\sfa\dsfb}
+L_\ssmD D  
+\frac{1}{4}(L^{\sfa\sfb}J_{\sfa\sfb}+L^{\dsfa\dsfb}J_{\dsfa\dsfb})
 + L^i{}_j J^j{}_i
\nonumber\\
&+&
 L_{{\ssmQ} i}^\sfa Q_\sfa^i
-L_{\ssmQ}^{{\dsfa}i} Q_{{{\dsfa}}i}
+L_{\ssmS}^{\sfa i}S_{\sfa i}
-L_{{\ssmS} i}^{{\dsfa}}S_{{\dsfa}}^i\,,
\end{eqnarray}
where $Q$ and $S$ are Poincar\'e and conformal supercharges respectively,
while $L_\ssmQ$ and $L_\ssmS$ are appropriate fermionic Cartan 1-forms
(for details see Appendix B).
The kinetic term and WZ three form take then the form

\begin{equation}\label{lkin6}
{\cal L}_{kin}
=-\frac{1}{2}\sqrt{g}g^{\vm\vn}
\Bigl(\hat{L}_\vm^{\sfa\dsfb}\hat{L}_{\vn\sfa\dsfb}
+L_{\ssmD\vm}L_{\ssmD\vn}
+L_\vm^\apr L_\vn^\apr\Bigr) \ ,
\end{equation}
\be\label{wzdec}
{\cal H} = 
{\cal H}_{AdS_5}^q+ {\cal H}_{S^5}^q -h.c. \ ,
\ee
where $AdS_5$ and $S^5$ contributions in WZ tree form are given by

\begin{eqnarray}
f{\cal H}_{AdS_5}^q
&=&
2{\rm i}\hat{L}^{\sfa\dsfb}
L_{{\ssmS} \sfa}^i\Csp_{ij}L_{{\ssmQ} {\dsfb}}^j
+L_\ssmD
(\frac{1}{2}L_{\ssmS}^{\sfa i}\Csp_{ij}L_{{\ssmS} \sfa}^j
+L_{\ssmQ}^{{\dsfa}i}\Csp_{ij}L_{{\ssmQ} {\dsfa}}^j)\,,
\\
\label{wzs5}
f{\cal H}_{S^5}^q
&=&
L_{\ssmS}^{\sfa i}(\Csp L)_{ij}L_{\ssmS \sfa}^j
-2L_{\ssmQ}^{{\dsfa}i}(\Csp L)_{ij}L_{\ssmQ \dsfa}^j\,.
\end{eqnarray}
We use the notation

\be\label{clij}
(C^\prime L)_{ij} \equiv C_{ik}^\prime L^k{}_j\,,
\ee
where $C_{ij}^\prime$ is a charge conjugation matrix of $SO(5)$
Dirac $\gamma$ matrices, 
$\gamma_\apr^T = C^\prime \gamma_\apr C^{\prime-1}$. Note that $L^\apr$ 
is expressible in terms of $L^i{}_j$ due to last relation in (\ref{lllll}).
The formulas (\ref{cf4})-(\ref{wzs5}) give description of 
$AdS_5\times S^5$ superstring action in conformal algebra notation. 
They are conformal counterpart of $so(4,1)\oplus so(5)$ representation 
given in (\ref{lkin2})-(\ref{cos55}).

\newsection{$psu(2,2|4)$ superalgebra and Cartan forms in \\
 $sl(2)\oplus su(4)$ basis}

To find superstring action in $sl(2)\oplus su(4)$ basis we need 
commutation relations and Cartan 1-forms in this basis. 
In this basis bosonic part of $psu(2,2|4)$ superalgebra consists 
of Poincar\'e translations $P_{\sfa\dsfb}$, conformal boosts 
$K_{\sfa\dsfb}$, dilatation $D$, Lorentz algebra generators 
$J_{\sfa\sfb}$, $J_{\dsfa\dsfb}$ and $su(4)$ algebra generators 
$J^i{}_j$. We adopt the following commutation relations between 
bosonic generators

\be
[P_{\sfa\dsfb},K_{\sfc\dsfd}]
=\epsilon_{\sfa\sfc}\epsilon_{\dsfb\dsfd}D
+\frac{1}{2}(\epsilon_{\sfa\sfc}J_{\dsfb\dsfd}
+\epsilon_{\dsfb\dsfd}J_{\sfa\sfc})\,,
\ee

\be\label{jjcom}
[J_{\sfa\sfb},J_{\sfc\sfd}] 
=\epsilon_{\sfb\sfc} J_{\sfa\sfd}+ 3 \hbox{ terms }\,,
\qquad
[J^i{}_j, J^k{}_n] = \delta_j^k J^i{}_n - \delta^i_n J^k{}_j\,,
\ee

\be
[D,P_{\sfa\dsfb}]=-P_{\sfa\dsfb}\,,
\qquad
[D,K_{\sfa\dsfb}]=K_{\sfa\dsfb}\,,
\ee

\be
[P_{\sfa\dsfb},J_{\sfc\sfd}] = \epsilon_{\sfa\sfc}P_{\sfd\dsfb}
+\epsilon_{\sfa\sfd}P_{\sfc\dsfb}\,,
\qquad
[K_{\sfa\dsfb},J_{\sfc\sfd}] = \epsilon_{\sfa\sfc}K_{\sfd\dsfb}
+\epsilon_{\sfa\sfd}K_{\sfc\dsfb}\,.
\ee
Note that because $J_{\sfa\sfb}$ is symmetric in $\sfa,\sfb$ 
the remaining three terms in the first commutator in (\ref{jjcom}) 
are obtainable by symmetrization in $\sfa,\sfb$ and $\sfc,\sfd$.
Fermionic part of $psu(2,2|4)$ superalgebra consists of 
Poincar\'e supergenerators $Q_\sfa^i$, $Q_{\dsfa i}$ and 
conformal supercharges
$S_{\sfa i}$, $S_\dsfa^i$. Commutation relations between 
bosonic and fermionic generators take the form

\be
[J_{\sfa\sfb},Q^i_\sfc] = \epsilon_{\sfb\sfc}Q^i_\sfa
+\epsilon_{\sfa\sfc}Q^i_\sfb\,,
\qquad
[J^i{}_j, Q^k_\sfa] =\delta_j^k Q^i_\sfa 
-\frac{1}{4}\delta_j^i Q^k_\sfa\,,
\ee

\be
[J_{\sfa\sfb},S_{\sfc i}] = \epsilon_{\sfb\sfc} S_{\sfa i}
+\epsilon_{\sfa\sfc} S_{\sfb i}\,,
\qquad
[J^i{}_j, S_{\sfa k}] = -\delta_k^i S_{\sfa j} 
+\frac{1}{4}\delta_j^i S_{\sfa k}\,,
\ee

\be
[D,Q^i_\sfa] = -\frac{1}{2}Q^i_\sfa\,,
\quad
[D,S_{\sfa i}]=\frac{1}{2} S_{\sfa i}\,,
\ee

\be
[S_{\sfa i},P_{\sfb{\dsfc}}]
=-{\rm i}\epsilon_{\sfa\sfb}Q_{\dsfc i}\,,
\qquad
[Q_{\sfa}^i,K_{\sfb \dsfc}]
={\rm i}\epsilon_{\sfa\sfb}S_{\dsfc}^i\,.
\ee
Anticommutation relations between fermionic generators take the form

\be
\{Q_{\sfa}^i,Q_{\dsfb j}\}
={\rm i}P_{\sfa \dsfb }\delta_j^i\,,
\qquad
\{S_{\sfa i},S_\dsfb^j\}
=-{\rm i}K_{\sfa{\dsfb}}\delta_i^j\,,
\ee

\be
\{Q_\sfa^i,S_{\sfb j}\}
=(\frac{1}{2}\epsilon_{\sfa\sfb}D+\frac{1}{2}J_{\sfa\sfb})\delta^i_j
+\epsilon_{\sfa\sfb}J^i{}_j\,.
\ee
Remaining nontrivial (anti)commutators can be obtained from the 
above ones via Hermitean conjugation rule which take the following form

\be
P_{\sfa \dsfb}^\dagger = - P_{\sfb \dsfa}\,,
\qquad
K_{\sfa \dsfb}^\dagger = - K_{\sfb \dsfa}\,,
\quad
J_{\sfa\sfb}^\dagger = J_{\dsfa\dsfb}\,,
\quad
(J^i{}_j)^\dagger = J^j{}_i\,,
\quad
D^\dagger = - D\,.
\ee

\be
(Q^i_\sfa)^\dagger = -Q_{\dsfa i}\,,
\quad
(Q^{\sfa i})^\dagger = Q^\dsfa_i\,,
\quad
(S_{\sfa i})^\dagger = -S^i_\dsfa\,,
\quad
(S^{\sfa}_i)^\dagger = S^{\dsfa i}\,.
\ee
The left invariant Cartan 1-forms in $sl(2)\oplus su(4)$ 
basis are defined by relation  (\ref{cf4}).
They satisfy the Maurer-Cartan equations
implied by the structure of $psu(2,2|4)$ superalgebra

\begin{eqnarray}
\label{maurer1}
dL_{\ssmP}^{\sfa\dsfb} 
&= & L_\ssmD\wedge L_{\ssmP}^{\sfa\dsfb}
-\frac{1}{2}(L^\sfa{}_\sfc\wedge L_\ssmP^{\sfc \dsfb}
+L^\dsfb{}_\dsfc\wedge L_\ssmP^{\sfa\dsfc})
-{\rm i}L_{\ssmQ i}^\sfa\wedge L_\ssmQ^{\dsfb i}\,,
\\
dL_{\ssmK}^{\sfa\dsfb} 
&=&
 -L_\ssmD\wedge L_{\ssmK}^{\sfa\dsfb}
-\frac{1}{2}(L^\sfa{}_\sfc\wedge L_\ssmK^{\sfc \dsfb}
+L^\dsfb{}_\dsfc\wedge L_\ssmK^{\sfa\dsfc})
+{\rm i}L_\ssmS^{\sfa i}\wedge L_{\ssmS i}^\dsfb\,,
\\
dL_\ssmD 
&=&
 -L_\ssmP^{\sfa\dsfb} \wedge L_{\ssmK \sfa\dsfb}
-\frac{1}{2}L_{\ssmQ i}^\sfa \wedge L_{\ssmS \sfa}^i
-\frac{1}{2}L_{\ssmQ }^{\dsfa i} \wedge L_{\ssmS \dsfa i}\,,
\\
\label{dlij}
dL^i{}_j 
&=&
 L^i{}_n \wedge L^n{}_j
-L_{\ssmQ j}^\sfa \wedge L_{\ssmS \sfa}^i
+L_{\ssmQ }^{\dsfa i} \wedge L_{\ssmS \dsfa j}-\hbox{Tr in } i,j\ \ ,
\\
dL_{\ssmQ}^{\dsfa i} 
&=&\frac{1}{2}L_\ssmD \wedge L_{\ssmQ }^{\dsfa i}
-\frac{1}{2}L^\dsfa{}_\dsfb \wedge L_{\ssmQ }^{\dsfb i}
+L^i{}_j \wedge L_{\ssmQ }^{\dsfa j}
+{\rm i}L_\ssmP^{\sfb\dsfa}\wedge L_{\ssmS \sfb}^i\,,
\\
dL_{\ssmQ i}^\sfa 
&=&
\frac{1}{2}L_\ssmD \wedge L_{\ssmQ i}^\sfa 
-\frac{1}{2}L^\sfa{}_\sfb \wedge L_{\ssmQ i}^\sfb
-L^j{}_i \wedge L_{\ssmQ j}^\sfa  
+{\rm i}L_\ssmP^{\sfa\dsfb}\wedge L_{\ssmS \dsfb i}\,,
\\
dL_{\ssmS }^{\sfa i} 
&=&
-\frac{1}{2}L_\ssmD \wedge L_{\ssmS}^{\sfa i} 
-\frac{1}{2}L^\sfa{}_\sfb \wedge L_{\ssmS }^{\sfb i}
+L^i{}_j \wedge L_{\ssmS }^{\sfa j}
-{\rm i}L_\ssmK^{\sfa\dsfb}\wedge L_{\ssmQ \dsfb }^i\,,
\\
\label{maurer10}
dL_{\ssmS i}^{\dsfa} 
&=&
-\frac{1}{2}L_\ssmD \wedge L_{\ssmS i}^{\dsfa } 
-\frac{1}{2}L^\dsfa{}_\dsfb \wedge L_{\ssmS i}^{\dsfb }
-L^j{}_i \wedge L_{\ssmS j}^{\dsfa }
-{\rm i}L_\ssmK^{\sfb\dsfa}\wedge L_{\ssmQ \sfb i}\,,
\end{eqnarray}
where the Tr in (\ref{dlij}) respects the condition $dL^i{}_i=0$.
There are Maurer-Cartan equations for $dL^{\sfa\sfb}$ 
and $dL^{\dsfa\dsfb}$ but 
we do not need them in what follows. 
Hermitean conjugation rules for Cartan 1-forms
take the form\footnote{
For fermionic coordinates we assume the convention 
$(\theta_1 \theta_2)^\dagger 
= \theta_2^\dagger \theta_1^\dagger$, 
$\theta_1\theta_2 =-\theta_2\theta_1$,
while for fermionic Cartan 1-form we adopt 
$(L_1 \wedge L_2)^\dagger = -L_2^\dagger \wedge L_1^\dagger$, 
$L_1\wedge L_2  = L_2\wedge L_1$.}

\be
L_{\ssmP,\ssmK}^{\sfa\dsfb*} = L_{\ssmP,\ssmK}^{\sfb\dsfa}\,,
\qquad
L_\ssmD^* = L_\ssmD\,,
\qquad
L^i{}_j^* = -L^j{}_i\,,
\qquad
L^{\sfa\sfb*} = -L^{\dsfa\dsfb}\,,
\ee
\be\label{carher}
(L_{\ssmQ i}^\sfa)^\dagger = L_{\ssmQ}^{\dsfa i}\,,
\quad
(L_{\ssmQ \sfa i})^\dagger = -L_{\ssmQ\dsfa}^i\,,
\quad
(L_\ssmS^{\sfa i})^\dagger = L_{\ssmS i}^\dsfa\,,
\quad
(L_{\ssmS \sfa}^i)^\dagger = -L_{\ssmS\dsfa i}\,.
\ee
The above relations are valid for arbitrary parametrization 
of supercoset space. 
To represent Cartan 1-forms in terms of bosonic and fermionic 
superstring coordinate fields
we fix the coset representative to be

\begin{equation}\label{G2}
G= g_{x,\theta}\  g_\eta \ g_y \ g_\phi \ ,
\end{equation}
where
\be
 g_{x,\theta} 
 =  \exp(x^{\sfa\dsfb}P_{\sfa\dsfb} +\theta_i^\sfa Q_\sfa^i
-\theta^{\dsfa i} Q_{{\dsfa }i}) \ ,
\qquad
 g_\eta  =  \exp(\eta^{\sfa i}S_{\sfa i}-\eta_i^{\dsfa }S_{\dsfa }^i)
 \ , 
\ee
and $g_\phi$ and $g_y$ depend on $AdS_5$ radial coordinate 
$\phi$ and $S^5$ coordinates $y^\apr$ respectively\footnote{Splitting 
fermionic coordinates in $\theta$ and $\eta$
was introduced in \cite{met1,met2}  in the study of 
light-cone gauge dynamics of  superparticle in  $AdS_5\times
S^5$.  Light-cone gauge superstring action in  $AdS_5\times S^5$
written in terms of these coordinates  was found in \cite{MT2}. 
In the context of brane dynamics these coordinates were discussed 
in \cite{pst}.}

\be\label{s5cos}
g_y = \exp(y^i{}_jJ^j{}_i)\,,
\qquad
g_\phi = \exp(\phi D)\,.
\ee
The $x^{\sfa\dsfb}$ and $y^i{}_j$ are expressible in terms of 
the coordinates along boundary directions $x^a$
and $S^5$ coordinates as follows

\be\label{yijya}
x^{\sfa\dsfb} = \frac{1}{\sqrt{2}}(\sigma^a)^{\sfa\dsfb } x^a\,,
\qquad
y^i{}_j = \frac{\rm i}{2}y^\apr (\gamma^\apr)^i{}_j\,,
\qquad
x^{\sfa\dsfb *} = x^{\sfb\dsfa } \,,
\quad
y^i{}_j^* = - y^j{}_i\,.
\ee
{}From these definitions it follows that $x^{\sfa\dsfb}$ 
and fermionic coordinates $\theta$, $\eta$ transform in appropriate 
representations of  $sl(2)\oplus su(4)$ algebra.
Plugging $G$ (\ref{G2}) in (\ref{cf4}) and after relative 
straightforward calculation we get the Cartan 1-forms 

\be\label{lpcar1}
L_\ssmP^{\sfa\dsfb}
=e^{\phi}{\sf L}_\ssmP^{\sfa\dsfb}\,,
\qquad
L_\ssmK^{\sfa\dsfb}
=e^{-\phi}{\sf L}_\ssmK^{\sfa\dsfb}\,,
\ee

\begin{eqnarray}
L_\ssmD & = & d\phi +\frac{1}{2}(\eta^{\sfa i} d\theta_{\sfa i}
+\eta^{\dsfa}_i d\theta_\dsfa^i)\,,
\\
\label{lijcar}
L^i{}_j &=&(dUU^{-1})^i{}_j
+\tilde{d\theta}_j^\sfa \tilde{\eta}_\sfa^i 
-\tilde{\eta}_j^\dsfa \tilde{d\theta}_\dsfa^i
-{\rm i}{\sf L}_\ssmP^{\sfa\dsfb}
\tilde{\eta}_{\dsfb j}\tilde{\eta}_\sfa^i
-\hbox{ Tr in } i,j \ \,,
\end{eqnarray}

\be\label{lllq0}
L_\ssmQ^{\dsfa i}=e^{\phi/2}U^i{}_j{\sf L}_\ssmQ^{\dsfa j}\,,
\qquad
L_{\ssmQ i}^{\sfa} 
={\sf L}_{\ssmQ j}^{\sfa} (U^{-1})^j{}_i e^{\phi/2}\,,
\ee

\be\label{llls0}
L_\ssmS^{\dsfa i}=e^{-\phi/2}U^i{}_j{\sf L}_\ssmS^{\dsfa j}\,,
\qquad
L_{\ssmS i}^{\sfa} 
={\sf L}_{\ssmS j}^{\sfa} (U^{-1})^j{}_i e^{-\phi/2}\,,
\ee
where the Tr in (\ref{lijcar}) respects the condition $L^i{}_i=0$ and 
we use the notation

\begin{eqnarray}
{\sf L}_{\ssmP}^{\sfa\dsfb}
&=&
dx^{\sfa\dsfb}-\frac{\rm i}{2}(\theta_i^\sfa d\theta^{\dsfb i}
+\theta^{\dsfb i}d\theta^\sfa_i)\,,
\\
\label{lllk1}
{\sf L}_\ssmK^{\sfa\dsfb}
&=&
 \frac{\rm i}{2}\eta^{\sfa i}d\eta_i^\dsfb
+\frac{\rm i}{12}(d\theta_i^\sfc
-\frac{\rm i}{4}{\sf L}_{\ssmP}^{\sfc\dsfd}\eta_{\dsfd i})
(5\eta_{\sfc}^j\eta^{\sfa i}
-\eta_{\sfc}^i\eta^{\sfa j}
-\eta^{\sfd i}\eta_\sfd^j\delta_\sfc^\sfa)\eta_j^\dsfb
+h.c.\,,
\\
\label{lllq1}
{\sf L}_\ssmQ^{\dsfa i}
&=&
d\theta^{\dsfa i} -{\rm i}\eta_\sfb^i {\sf L}_{\ssmP}^{\sfb\dsfa}\,,
\qquad
{\sf L}_{\ssmQ i}^{\sfa}
= d\theta_i^\sfa  -{\rm i}{\sf L}_{\ssmP}^{\sfa\dsfb}\eta_{\dsfb i}\,,
\\
{\sf L}_\ssmS^{\sfa i}
&=&d\eta^{\sfa i}+ \frac{1}{4}(d\theta_j^\sfc  
-\frac{\rm i}{3}{\sf L}_{\ssmP}^{\sfc\dsfd}\eta_{\dsfd j})
(\eta_{\sfc}^j\eta^{\sfa i}
-3\eta_{\sfc}^i\eta^{\sfa j}
+\eta^{\sfd j}\eta_\sfd^i\delta_\sfc^\sfa)
\nonumber
\\
\label{llls1}
&+&\frac{1}{2}(d\theta^{\dsfc i} 
-\frac{\rm i}{3}\eta_\sfb^i {\sf L}_{\ssmP}^{\sfb\dsfc})
\eta_{\dsfc j}\eta^{\sfa j}\,.
\end{eqnarray}
The $\tilde{\theta}$ and $\tilde{d\theta}$ are defined by
\begin{equation}\label{tilthe}
\tilde{\theta}^{\dsfa i} \equiv  U^i{}_j \theta^{\dsfa j}\,,
\qquad
\tilde{\theta}_{\sfa i} \equiv  \theta_{\sfa j} (U^{-1})^j{}_i  \ ,
\end{equation}
\begin{equation}\label{tildthe}
\tilde{d\theta}{}^{\dsfa i} \equiv U^i{}_j d\theta^{\dsfa j}\,,
\qquad
\tilde{d\theta}_{\sfa i} \equiv  d\theta_{\sfa j} (U^{-1})^j{}_i      \  ,
\end{equation}
and similar ones for $\eta$.
Hermitean conjugation in 
(\ref{lllk1}) should be supplemented by $\sfa \leftrightarrow \sfb$.
The matrix $U\in SU(4) $ is defined  in terms of
the $S^5$ coordinates  $y^i_{\ j}$  (or $y^\apr$, see (\ref{yijya}))
by $U^i{}_j = (e^y)^i{}_j$. It can be   written  explicitly as
\be
U=\cos{|y|\over 2 } +{\rm i}\gamma^\apr n^\apr \sin{|y|\over 2 }
\,,
\qquad
|y|\equiv\sqrt{y^\apr y^\apr}\,,
\quad
n^\apr \equiv  \frac{y^\apr}{|y|}  \ .   \la{uuuu}
\ee
Because fermionic coordinates defined by (\ref{G2}) 
trasform covariantly the 1-forms defined in (\ref{lllq1}),(\ref{llls1}) 
also transform covariantly under $SU(4)$. On the other 
hand because the  
matrix $U^i{}_j$ does not transform covariantly under $SU(4)$ 
group the Cartan 1-forms defined by (\ref{lllq0}),(\ref{llls0}) 
also do not transform covariantly under this group. Below we will 
demonstrate that superstring action can be expressed entirely in 
terms of $\eta$ and 1-forms given by 
(\ref{lllq1}),(\ref{llls1}) and this explains why the superstring 
action can be put into manifestly $SU(4)$ covariant form.

\newsection{Manifestly $SL(2,C)\times SU(4)$ invariant form of \\
superstring action}

Plugging the expressions for Cartan 1-froms (\ref{lpcar1})-(\ref{llls0}) 
into expressions for kinetic and WZ terms (\ref{lkin6}),(\ref{wzdec}) 
we can represent the superstring Lagrangian in terms of string 
coordinate fields.
Let us start with kinetic term ${\cal L}_{kin}$. Contribution of 
$AdS_5$ part in ${\cal L}_{kin}$ has manifest $SL(2,C)\times SU(4)$ 
invariance from the very beginning. In order to transform contribution 
of $S^5$ part into desired form we notice that $L^\apr$ given by 
(\ref{lllll}) and (\ref{lijcar}) can be cast into the form
 
\be\label{lapr}
L^\apr = e_\sca^\apr(dy^\sca 
-d\theta_i^\sfa ({\cal V}^\sca)^i{}_j \eta_\sfa^j
+\eta_i^\dsfa ({\cal V}^\sca)^i{}_j d\theta_\dsfa^j
+{\rm i}{\sf L}_\ssmP^{\sfa\dsfb}
\eta_{\dsfb i}({\cal V}^\sca)^i{}_j\eta_\sfa^j)\,,
\ee
where $e_\sca^\apr$ are components of $S^5$ vielbein 
$e^\apr =e_\sca^\apr dy^\sca$: 

\begin{equation}\label{gab}
G_{\sca\scb}= e_\sca^\apr e_\scb^\apr\,,   \ \
\qquad
e_\sca^\apr=\frac{\sin |y|}{|y|}(\delta_\sca^\apr - n_\sca n^\apr)
+n_\sca n^\apr\,,
\end{equation}
 and $({\cal V}^\sca)^i{}_j$ are  the components of
the Killing vectors $({\cal V}^\sca)^i{}_j  \partial_{y^\sca}$ of $S^5$ \ 
($\partial_{y^\sca}  = \partial/\partial  y^\sca$).
The $G_{\sca\scb}$ is metric tensor on $S^5$ in 
the coordinates $y^\sca$ defined by coset representative 
$g_y$ (\ref{s5cos}),(\ref{yijya}).
The $({\cal V}^\sca)^i{}_j\partial_{y^\sca}$ satisfy  the  
$so(6)\simeq su(4)$
commutation relations  (\ref{jjcom}) and  may be written as

\begin{equation}
({\cal V}^\sca)^i{}_j\partial_{y^\sca}
=\frac{1}{4}(\gamma^{\apr\bpr})^i{}_jV^{\apr\bpr}
+\frac{\rm i}{2}(\gamma^\apr)^i{}_j V^\apr      \ ,
\end{equation}
where  $V^\apr$ and $V^{\apr\bpr}$
correspond to the  5 translations and  $SO(5)$
rotations respectively and  are given by 

\begin{eqnarray}
V^\apr
=\Bigl[|y|\cot |y| (\delta^{\apr\sca}-n^\apr n^\sca)
+n^\apr n^\sca\Bigr]\partial_{y^\sca}  \ ,
\qquad
V^{\apr\bpr}=y^\apr \partial_{y^\bpr}-y^\bpr \partial_{y^\apr}  \ .
\end{eqnarray}
They satisfy the $so(6)$ algebra commutation relations given 
in (\ref{cmr1}),(\ref{cmr2}).
Here $\delta^{\apr\sca}$ is  Kronecker delta symbol and we
use the conventions: $y^\sca=\delta_\apr^\sca y^\apr$,\
$n^\sca=\delta_\apr^\sca n^\apr$, $n^\sca=n_\sca$.
Note that while deriving (\ref{lapr}) we use the relation 

\be
(U^\dagger \gamma^\apr U)^i{}_j=-2{\rm i}e_\sca^\apr({\cal V}^\sca)^i{}_j\,.
\ee
The $L^\apr L^\apr$ 
can be put into the  manifestly $SU(4)$ 
invariant form by  changing the coordinates from
$y^\apr$ to the  $6d$ unit vector $u^M$\ ($M=1,...,6$):

\be\label{uvec}
u^\apr = n^\apr \sin|y|\,, 
\qquad
u^6 =  \cos|y|\,,
\qquad
u^M u^M = 1\,, 
\ee
and noticing the following important relation

\be\label{laprdum}
L^\apr L^\apr = Du^M Du^M\,,
\ee
where
\be\label{dum}
Du^M =du^M
-d\theta_i^\sfa (R^M)^i{}_j \eta_\sfa^j
+\eta_i^\dsfa (R^M)^i{}_j d\theta_\dsfa^j
+{\rm i}{\sf L}_\ssmP^{\sfa\dsfb}\eta_{\dsfb i}(R^M)^i{}_j\eta_\sfa^j\,,
\ee
\be
R^M \equiv - \frac{1}{2}\rho^{MN}u^N\,.
\ee
To derive (\ref{laprdum}) we used the identities 

\be
G_{\sca\scb}(\lambda {\cal V}^\sca \theta)
(\vartheta {\cal V}^\scb \eta)
=(\lambda R^M \theta)(\vartheta R^M\eta)\,,
\qquad
G_{\sca\scb}(\lambda {\cal V}^\sca \theta) dy^\scb
=(\lambda R^M \theta)du^M\,,
\ee
where expressions like $(\lambda {\cal V}^\sca \theta)$ stand for
$\lambda_i ({\cal V}^\sca)^i{}_j \theta^j$.
Making use of these relations we get the following manifest 
$SL(2,C)\times SU(4)$ invariant representation for kinetic term

\be\label{kin7}
{\cal L}_{kin} = -\frac{1}{2}\sqrt{g}g^{\vm\vn}(\hat{L}^{\sfa\dsfb}_\vm
\hat{L}_{\vn\sfa\dsfb} + D_\vm \phi D_\vn\phi +D_\vm u^M D_\vn u^M)\,,
\ee
where

\be
D\phi \equiv    d\phi +\frac{1}{2}(\eta^{\sfa i} d\theta_{\sfa i}
+\eta^{\dsfa}_i d\theta_\dsfa^i)\,.
\ee
Note that by  changing the coordinates from
$\phi$, $u^M$ to $Z^M$ defined by

\be\label{zdef}
Z^M = e^{-\phi } u^M\,,
\ee
we can cast the kinetic term (\ref{kin7}) into another manifest 
$SL(2,C)\times SU(4)$ invariant form given in (\ref{lag0kin}).

We proceed now to the WZ three form.
To transform  the WZ term into desired form 
we  use the standard  trick of rescaling
$\eta\rightarrow \eta_t \equiv t\eta$ (note 
that we do not rescale $\theta$)

\be
{\cal H} =  {\cal H}_t(t=1)\,,
\qquad
{\cal H}_t = {\cal H} (\eta_t) \ .
\ee
Then  one has the obvious relation 

\be\label{hdec9}
{\cal H} = {\cal H}_{t=0} + \int_0^1 dt \, \partial_t {\cal H}_t\,.
\ee
Note that shifting $\eta\rightarrow t\eta$ in coset representative 
(\ref{G2}) and then setting $t=0$ corresponds to chosing $S$ gauge, i.e. 
${\cal H}_{t=0}$ is nothing but WZ three form in $S$ gauge. 
Its contribution to ${\cal L}_{WZ}$ has been already 
find in \cite{pes,kr,MT2} (note we use convention of \cite{MT2}, where 
the details of deriving 
${\cal H}_{t=0}$ may be found) and is given by

\be\label{lwz8}
f {\cal L}_{WZ}|_{t=0}
= {\rm i}e^\phi d \theta^{\dsfa i} y_{ij} d \theta_\dsfa^j + h.c.\,,
\ee
where

\be
y_{ij} \equiv \rho_{ij}^M u^M\,.
\ee
To derive (\ref{lwz8}) we used the relation
$U^T C^\prime = C^\prime U$
and the representation for charge 
conjugation $C_{ij}^\prime$ matrix and  $SO(5)$ Dirac $\gamma$ 
matrices
given by (\ref{Cg}) which together 
with (\ref{uuuu}) gives the following important relation

\be\label{cijyij}
(C^\prime U^2)_{ij} = y_{ij}\,.
\qquad
\ee
The relation (\ref{cijyij}) 
expresses the fact that though neither $C^\prime$ matrix or $U$ matrix 
do not transform covariantly under $SU(4)$ their special combination 
(\ref{cijyij}) transforms covariantly under $SU(4)$.
Now we consider contribution of the second term in (\ref{hdec9}).
Making shift $\eta\rightarrow t\eta$ in (\ref{G2}) and using 
(\ref{cf4}) we get 
the following equations

\be\label{dt1}
\partial_t L_\ssmK^{\sfa\dsfb} 
= -{\rm i} L_{\ssmS i}^\dsfb \tilde{\tilde{\eta}}{}^{\sfa i}
-{\rm i} L_{\ssmS}^{\sfa i} \tilde{\tilde{\eta}}{}^{\dsfb }_i\,,
\ee
\be\label{dt6}
\partial_t L_\ssmD
= \frac{1}{2} L_{\ssmQ i}^\sfa \tilde{\tilde{\eta}}{}_{\sfa}^i
+h.c.\,,
\qquad
\partial_t L^i{}_j  
=L_{\ssmQ j}^\sfa \tilde{\tilde{\eta}}{}_{\sfa}^i
- \frac{1}{4}\delta^i_j L_{\ssmQ n}^\sfa \tilde{\tilde{\eta}}{}_{\sfa}^n
-h.c.\,,
\ee
\be\label{dt7}
\partial_t L_\ssmQ^{\dsfa i} 
= {\rm i} L_{\ssmP}^{\sfb \dsfa} \tilde{\tilde{\eta}}{}_\sfb^i\,,
\qquad
\partial_t L_\ssmS^{\sfa i} 
= d\tilde{\tilde{\eta}}{}^{\sfa i}
+\frac{1}{2}L_\ssmD \tilde{\tilde{\eta}}{}^{\sfa i}
+\frac{1}{2}L^\sfa{}_\sfb \tilde{\tilde{\eta}}{}^{\sfb i}
-L^i{}_j\tilde{\tilde{\eta}}{}^{\sfa j}\,,
\ee
where
\be
\tilde{\tilde{\eta}}{}^{\sfa i} 
\equiv e^{-\phi/2}U^i{}_j \eta^{\sfa j}\,,
\qquad
\tilde{\tilde{\eta}}{}_{\sfa i} 
\equiv \eta_{\dsfa j}(U^{-1})^j{}_i e^{-\phi/2}\,,
\ee
and we assume that in expressions for Cartan 1-forms 
in (\ref{dt1})-(\ref{dt7})
all $\eta$ are shifted $\eta \rightarrow t\eta$.
Hermitean conjugation in $\partial_t L^i{}_j$ in (\ref{dt6})
should be supplemented by $i \leftrightarrow j$.
Now making use of (\ref{wzdec}), formulas (\ref{dt1})-(\ref{dt7}) 
and Maurer-Cartan equations (\ref{maurer1})-(\ref{maurer10}) we find 

\be\label{dth}
\partial_t (f{\cal H}^q)
=d\Bigl[ -2{\rm i}\hat{L}^{\sfa \dsfb}
\tilde{\tilde{\eta}}{}_\sfa^i C_{ij}^\prime L_{\ssmQ \dsfb}^j
-L_\ssmD \tilde{\tilde{\eta}}^{\sfa i}C_{ij}^\prime L_{\ssmS \sfa}^j
+\tilde{\tilde{\eta}}^{\sfa i}((C^\prime L)_{ij}
-(C^\prime L)_{ji})L_{\ssmS \sfa}^j\Bigr]\,,
\ee
where ${\cal H}^q \equiv {\cal H}_{AdS_5}^q+{\cal H}_{S^5}^q$.
Making use of formula (\ref{dth}) and the relation

\be
(U^T(C^\prime L)U)_{ij} - (i \leftrightarrow j)
= Du^M \rho_{ij}^M\,,
\ee
where $(C^\prime L)_{ij}$ is defined by (\ref{clij}) 
we get finally the desired manifest $SL(2,C)\times SU(4)$ 
representation for ${\cal L}_{WZ}$

\be
f{\cal L}_{WZ}
={\rm i}e^\phi d \theta^{\dsfa i} y_{ij} d \theta_\dsfa^j
+\int_0^1 \! dt [2\hat{L}^{\sfa \dsfb}
\eta_\sfa^i y_{ij}{\sf L}_{\ssmQ \dsfb}^j
-{\rm i}e^{-\phi}L_\ssmD \eta^{\sfa i} y_{ij} {\sf L}_{\ssmS \sfa}^j
+{\rm i}e^{-\phi}\eta^{\sfa i}Dy_{ij} {\sf L}_{\ssmS \sfa}^j\Bigr]
+h.c.
\ee
where $Dy_{ij}= \rho_{ij}^M Du^M$ and $Du^M$ is defined by 
(\ref{dum}). In terms of coordinates $Z^M$ defined by (\ref{zdef}) 
we can cast this ${\cal L}_{WZ}$ into the form given by (\ref{lwz1}).

\section*{Acknowledgments}

This  work  was  supported  by
the DOE grant DE-FG02-91ER-40690, by the INTAS project 991590,
and by the RFBR Grant No.99-02-17916.

\setcounter{section}{0}
\setcounter{subsection}{0}

\appendix{Notation}
\label{not}

In the main part of the paper we 
 use the following  conventions for the indices:
\begin{eqnarray*}
\aA,\aB,\aC=0,\ldots, 4 && \qquad  so(4,1)\  \hbox{vector
indices ($AdS_5$ tangent space indices)}
\\
\apr,\bpr,\cpr=1,\ldots, 5 && \qquad  so(5)\  \hbox{ vector
indices ($S^5$ tangent space indices) }
\\
a, b, c =0,\ldots, 3 & &\qquad
\hbox{boundary Minkowski space indices}
\\
{\cal A}, {\cal B}, {\cal C}=1,\ldots, 5 & &\qquad
\hbox{$S^5$  coordinate space indices}
\\
M,N,K,L =1,\ldots, 6 
&& \qquad so(6)\  \hbox{vector indices}
\\
i,j,k,n =1,\ldots, 4 && \qquad su(4)\  \hbox{vector indices}
\\
\alpha,\beta,\gamma = 1,\ldots, 4 
&& \qquad  so(4,1) \  \hbox{ spinor indices }
\\
\sfa,\sfb,\sfc;\  (\dsfa,\dsfb,\dsfc) = 1,2; \  (\dot{1},\dot{2}) 
&& \qquad  sl(2,C)\  \hbox{ spinor
indices }
\\
\mu,\nu = 0,1 
&&
\qquad
\hbox{world sheet coordinate indices}
\end{eqnarray*}
Note that we identify the $su(4)$ 
vector indices $i,j,k,n$ with $so(5)$ spinor indices.
We decompose $x^a$ into the light-cone and  2 complex coordinates:
$x^a= (x^+,x^-,x, \bar{x})$

\be
x^\pm\equiv \frac{1}{\sqrt{2}}(x^3\pm x^0)\,,
\qquad \ \ \
x,\bar{x} = \frac{1}{\sqrt{2}} (x^1 \pm  {\rm i} x^2)\,.
\end{equation}
We suppress the flat space metric tensor $\eta_{ab}=(-,+,+,+)$ 
in scalar products, i.e. 
$A^a B^a\equiv \eta_{ab}A^a B^b $.
The  world-sheet 
Levi-Civita $\epsilon^{\vm\vn}$   is defined with   $\epsilon^{01}=1$.

We use the following decomposition of $so(4,1)$
Dirac  $\gamma^\aA$ and  charge conjugation  $C_{\alpha\beta}$ 
matrices in the $sl(2)$ basis
 
\begin{equation}\label{gamdec}
(\gamma^a)^\alpha{}_\beta
=\left(
\begin{array}{cc}
0 & (\sigma^a)^{\sfa\dsfb  }
\\
\bar{\sigma}^a_{\dsfa \sfb}  & 0
\end{array}
\right)\,,
\qquad
\gamma^{4}
=\left(
\begin{array}{cc}
1 & 0
\\
0  & -1
\end{array}
\right)\,,
\qquad
C_{\alpha\beta}
=\left(
\begin{array}{cc}
\epsilon_{\sfa\sfb} & 0
\\
0  & \epsilon^{\dsfa \dsfb  }
\end{array}
\right)\,,
\end{equation}
where  the matrices $(\sigma^a)^{\sfa\dsfa}$,
$(\bar{\sigma}^a)_{\dsfa\sfa}$ are related to Pauli matrices
in the standard way
\be
(\sigma^a)^{\sfa\dsfa} =(1,\sigma^1,\sigma^2,\sigma^3)\,,
\qquad
(\bar{\sigma}^a)_{\dsfa\sfa} =(-1,\sigma^1,\sigma^2,\sigma^3)\, . 
\ee
Note that 
$\bar{\sigma}_{\dsfa\sfb}^a
=\sigma_{\sfb\dsfa}^a=\sigma_{\sfa\dsfb}^{a*}$ where
$\sigma_{\sfa\dsfa}^a\equiv
(\sigma^a)^{\sfb\dsfb}\epsilon_{\sfb\sfa}\epsilon_{\dsfb\dsfa}$.
Sometimes we use the relation
\be
(\sigma^a)^{\sfa\dsfa}(\sigma^a)^{\sfb\dsfb}
=2\epsilon^{\sfa\sfb}\epsilon^{\dsfa\dsfb}\,.
\ee
We use the following conventions for the $sl(2)$ indices: \ 
$
\epsilon_{12}=\epsilon^{12}=-\epsilon_{\dot{1}\dot{2}}
=-\epsilon^{\dot{1}\dot{2}}=1    \ ,
$
\begin{equation}\label{rlrul}
\psi^\sfa=\epsilon^{\sfa\sfb}\psi_\sfb  \ ,
\quad
\psi_\sfa=\psi^\sfb\epsilon_{\sfb\sfa}    \ ,
\qquad
\psi^{\dsfa }=\epsilon^{\dsfa \dsfb  }\psi_{\dsfb  }\ ,
\quad
\psi_{\dsfa }=\psi^{\dsfb  }\epsilon_{\dsfb  \dsfa } \ .
\end{equation}
Note that $(\epsilon^{\sfa\sfb})^* = -\epsilon^{\dsfa\dsfb}$.
We use the notation

\be
(\sigma^{ab})_{\sfa\sfb}= (\sigma^{ab})^\sfc{}_\sfb\epsilon_{\sfc\sfa}\ ,
\qquad
(\bar{\sigma}^{ab})_{\dsfa\dsfb}= 
(\bar{\sigma}^{ab})_\dsfa{}^\dsfc\epsilon_{\dsfc\dsfb}\ ,
\ee
where

\be
(\sigma^{ab})^\sfa{}_\sfb\equiv\frac{1}{2}
(\sigma^a)^{\sfa\dsfc}(\bar{\sigma}^b)_{\dsfc\sfb}
-(a\leftrightarrow b)   \ ,
\qquad
(\bar{\sigma}^{ab})_\dsfa{}^\dsfb\equiv\frac{1}{2}
(\bar{\sigma}^a)_{\dsfa\sfc}(\sigma^b)^{\sfc\dsfb}
-(a\leftrightarrow b)   \ .
\ee
Note that $(\sigma^{ab})_{\sfa\sfb} = (\sigma^{ab})_{\sfb\sfa}$,
$(\sigma^{ab})_{\sfa\sfb}^*=(\bar{\sigma}^{ab})_{\dsfa\dsfb}$.

The six  matrices   $\rho_{ij}^M$  represent 
the $SO(6)$   Dirac   matrices $\gamma^M$
in the  chiral representation, i.e. 
\be\label{usgam}
\gamma^M
=\left(\begin{array}{cc}
 0   & (\rho^M)^{ij} 
 \\
 \rho_{ij}^M & 0
 \end{array}
 \right)\,, \ee
 \be 
 (\rho^M)^{ik}\rho_{kj}^N + (\rho^N)^{ik}\rho_{kj}^M
 =2\delta^{MN}\delta_j^i\,,
 \qquad
 \rho_{ij}^M =- \rho_{ji}^M\,,
 \qquad (\rho^M)^{ij}\equiv  - (\rho_{ij}^{M})^* \ . 
 \ee
The $SO(5)$  Dirac  and  charge conjugation 
 matrices  can be expressed  in terms 
of the $\rho^M$ matrices as   follows 

\be\label{Cg}
(\gamma^\apr)^i{}_j = {\rm i}(\rho^\apr)^{ik}\rho_{kj}^6\,,\ \ 
\qquad
C_{ij}^\prime =\rho_{ij}^6 \  . 
\ee
The $\rho^M$ matrices satisfy the identities

\be
\rho_{ij}^M=\frac{1}{2}\epsilon_{ijkn}(\rho^M)^{kn}\,,
\qquad
\rho_{ij}^M(\rho^M)^{kn}
=2(\delta_i^n\delta_j^k-\delta_i^k\delta_j^n) \ . 
\la{iii}
\ee
The matrices $\rho^{MN}$  are defined by 

\be
(\rho^{MN})^i{}_j \equiv \frac{1}{2}(\rho^M)^{ik}\rho_{kj}^N
-(M\leftrightarrow N)\,, 
\la{rrr}
\ee
so that 

\be
(\rho^{MN})^i{}_j (\rho^{MN})^k{}_n= 2\delta^i_j\delta^k_n
 -8\delta^i_n\delta^k_j\,.
\ee

\appendix{Transformation of $psu(2,2|4)$ superalgebra from $so(4,1)\oplus
so(5)$ to $sl(2,C)\oplus su(4)$ basis}

We start with the commutation
relations of $psu(2,2|4)$ superalgebra in $so(4,1)\oplus so(5)$ basis
given in \ci{MT}

\begin{equation}\label{cmr1}
[\hat{P}_\aA,\hat{P}_\aB]=\hat{J}_{\aA\aB}\,,
\qquad
[P_\apr, P_\bpr]=-J_{\apr\bpr}\,,
\end{equation}
\begin{equation}\label{cmr2}
[\hat{J}^{\aA\aB},\hat{J}^{\aC\aE}]
=\eta^{\aB\aC}\hat{J}^{\aA\aE}+3 \hbox{ terms},
\qquad
[J^{\apr\bpr},J^{\cpr\epr}]=\eta^{\bpr\cpr}J^{\apr\epr}+3 \hbox{ terms} \
,
\end{equation}
\be
[Q_{_I},\hat{P}_\aA]
=-\frac{{\rm i}}{2}\epsilon_{_{IJ}}Q_{_J}\gamma_\aA \,,
\qquad
[Q_{_I},\hat{J}_{\aA\aB}]=-\frac{1}{2} Q_{_I}\gamma_{\aA\aB}\,,
\ee
\be
[Q_{_I},P_\apr]
=\frac{1}{2}\epsilon_{_{IJ}} Q_{_J}\gamma_\apr\,,
\qquad
[Q_{_I},J_{\apr\bpr}]=-\frac{1}{2}Q_{_I}\gamma_{\apr\bpr} \,,
\ee
\begin{eqnarray}
\{Q_{\alpha \alpr I}, Q_{\beta \bepr J}\}
&=&\delta_{_{IJ}}
[-2{\rm i}C_{\alpr\bepr}^\prime(C\gamma^\aA)_{\alpha\beta} \hat{P}_\aA
+2C_{\alpha\beta}(C^\prime\gamma^\apr)_{\alpr\bepr}P_\apr]
\nonumber\\
&+&\epsilon_{_{IJ}}
[C_{\alpr\bepr}^\prime(C\gamma^{\aA\aB})_{\alpha\beta} \hat{J}_{\aA\aB}
-C_{\alpha\beta}(C^\prime\gamma^{\apr\bpr})_{\alpr\bepr}J_{\apr\bpr}] \ .
\end{eqnarray}
where $\epsilon_{12}=-\epsilon_{21}=1$,
$\gamma_{AB}=\frac{1}{2}\gamma_A\gamma_B -(A \leftrightarrow B)$.
Unless otherwise  specified, we use the notation
$Q^I$ for $Q^{I\alpha\alpr}$ and $Q_I$ for $Q_{I\alpha\alpr}$, 
where 

\be\label{q1q2}
Q_{I\alpha\alpr}\equiv
Q^{J\beta\bepr}\delta_{JI}C_{\beta\alpha}C_{\bepr\alpr}^\prime\,.
\ee
Hermitean conjugation rules in this basis are
\be
\hat{P}_\aA^\dagger=- \hat{P}_\aA\,,\ \ \ \ \
P_\apr^\dagger= - P_\apr \ , \ \ \
\ \
\hat{J}_{\aA\aB}^\dagger=- \hat{J}_{\aA\aB}\,,\ \ \ \ \
\ \
J_{\apr\bpr}^\dagger=- J_{\apr\bpr}
\ ,\ee \be  \ \
(Q^{I\beta\alpr})^\dagger (\gamma^0)^\beta_\alpha
= - Q^{I\beta\bepr}C_{\beta\alpha}C_{\bepr\alpr}^\prime\ .
\ee
The transformation of the  bosonic generators into
the conformal algebra basis are given by formulas (\ref{dp4})
Let us describe transformation of fermionic generators.
First we introduce the new ``charged"  super-generators

\be
Q^q\equiv \frac{1}{\sqrt{2}}(Q^1+{\rm i}Q^2)   \ ,
\qquad  \ \ \
Q^{\bar{q}}
\equiv \frac{1}{\sqrt{2}}(Q^1-{\rm i}Q^2)
\ . \ee
We shall use the simplified notation
\begin{equation}\label{intsc2}
Q^{\alpha\alpr}\equiv -Q^{q\alpha\alpr}\,,
\qquad
Q_{\alpha\alpr}\equiv Q_{q\alpha\alpr}\,.
\end{equation}
Then   the non-vanishing values of
$\delta_{IJ}$  ($\epsilon_{IJ}$, $\epsilon_{12}=1$) become replaced by
$\delta_{q\bar{q}}=1$ ($\epsilon_{q\bar{q}}={\rm i}$) and
the Majorana condition  takes  the form  $
(Q^{\beta\alpr})^\dagger(\gamma^0)^\beta_\alpha
=Q_{\alpha\alpr}$. The relation (\ref{q1q2}) and (\ref{intsc2}) give

\be
Q_{\bar{q}\alpha i} = -Q^{\beta j}C_{\beta \alpha} C_{ji}^\prime\,.
\ee
We then decompose the supercharges in the $sl(2)\oplus su(4)$ basis
\be
Q^{\alpha i}
=\left(
\begin{array}{c}
2{\rm i}v^{-1}Q^{\sfa  i}
\\[7pt]
2vS_{\dsfa }^i
\end{array}\right)  \ ,
\qquad    \
Q_{\alpha i}=(2vS_{\sfa i}, -2{\rm i}v^{-1}Q_i^{\dsfa })   \ ,   \ \ \
\ \  v\equiv 2^{1/4} \ .
\ee
In terms of these  new supercharges  the commutation relations take the
form given in the Section 3.

Now let us consider  the fermionic  1-forms. They satisfy
hermitean conjugation rule
\be
(L^{I\beta\alpr})^\dagger (\gamma^0)^\beta_\alpha
= L^{I\beta\bepr}C_{\beta\alpha}C_{\bepr\alpr}^\prime.
\ee
and we use the notation
$L_{I\alpha\alpr}\equiv
L^{J\beta\bepr}\delta_{JI}C_{\beta\alpha}C_{\bepr\alpr}^\prime$.
Let us   define
\begin{equation}\label{lq}
L^q\equiv \frac{1}{\sqrt{2}}(L^1+{\rm i}L^2)   \ ,
\qquad  \ \
L^{\bar{q}}\equiv \frac{1}{\sqrt{2}}(L^1-{\rm i}L^2) \ ,
\end{equation}
and introduce the notation
$L^{\alpha i}= L^{q\alpha i}$, $L_{\alpha i}= L_{q\alpha i}$. 
These relations give

\be
L_{\alpha i} = L^{\bar{q}\beta j}C_{\beta \alpha} C_{ji}^\prime\,.
\ee
We use then the following decomposition into
$sl(2)\oplus su(4)$ Cartan 1-forms
\be
L^{\alpha i}
=\frac{1}{2} \left(
\begin{array}{c}
v^{-1}L_\ssmS^{\sfa  i}
\\[7pt]
{\rm i}v L_{\ssmQ\,\dsfa }^i
\end{array}\right) \ , \ \ \ \
\qquad
L_{\alpha i}=\frac{1}{2} (- {\rm i}v L_{\ssmQ\,\sfa i},\,
v^{-1}L_{\ssmS i}^{\dsfa })
\ . \ee
Hermitean conjugation rules for the new Cartan 1-forms 
in $sl(2)\oplus su(4)$ basis take then the form given in (\ref{carher}).
The above relations lead to the decomposition 
$L^{I\alpha i} Q_{I\alpha i} =
L^{\alpha i}Q_{\alpha i} - L_{\alpha i}Q^{\alpha i}$ which 
in terms of $sl(2)\oplus su(4)$ notation takes the form given 
in the second line in (\ref{cf4}).

\end{document}